\documentclass[graybox]{svmult}

\usepackage{type1cm}        
\usepackage{makeidx}         
\usepackage{graphicx}        
\usepackage{multicol}        
\usepackage[bottom]{footmisc}

\usepackage{newtxtext}       %
\usepackage{newtxmath}       

\usepackage{amsmath}
\usepackage{mathtools}
\usepackage{bbm}

\usepackage[colorlinks=true,
linkcolor=blue,
urlcolor=blue,
citecolor=blue,
bookmarks=true,
pdfborder={0 0 0}]{hyperref}

\newcommand{\En}[1]{\mathbb{E} \left( #1 \right)}

\DeclarePairedDelimiter{\ceil}{\lceil}{\rceil}

\makeindex             

\begin{document}

\title*{Social contagion on higher-order structures}

\author{Alain Barrat, 
Guilherme Ferraz de Arruda,
Iacopo Iacopini and
Yamir Moreno}
\institute{Alain Barrat \at 
Aix Marseille Univ, Universit\'e de Toulon, CNRS, CPT, Marseille, France, \email{alain.barrat@cpt.univ-mrs.fr}
\and 
Guilherme Ferraz de Arruda  \at ISI Foundation, Torino, Italy
\email{gui.f.arruda@gmail.com}
\and
Iacopo Iacopini \at 
Aix Marseille Univ, Universit\'e de Toulon, CNRS, CPT, Marseille, France, \email{iacopiniiacopo@gmail.com}
\and
Yamir Moreno \at 
Institute for Biocomputation and Physics of Complex Systems (BIFI), University of Zaragoza, 50018 Zaragoza, Spain
\email{yamir.moreno@gmail.com}
}
%
%
\maketitle

\abstract{In this Chapter, we discuss the effects of higher-order structures on SIS-like processes of social contagion. After a brief motivational introduction where we illustrate the standard SIS process on networks and the difference between simple and complex contagions, we introduce spreading processes on higher-order structures starting from the most general formulation on hypergraphs and then moving to several mean-field and heterogeneous mean-field approaches. The results highlight the rich phenomenology brought by taking into account higher-order contagion effects: both continuous and discontinuous transitions are observed, and critical mass effects emerge. We conclude with a short discussion on the theoretical results regarding the nature of the epidemic transition and the general need for data to validate these models.}

\section{Introduction}
\label{sec:Introduction}

The standard modeling and study of social or biological contagion processes in populations is based on two types of ingredients.
First, the evolution of the process within each 
individual is often described through compartmental models
\cite{anderson1992infectious,keeling2011modeling}, such that
each individual is at any time in one of several possible compartments or states. For instance, in the description of many infectious diseases, the
considered states include
susceptible (S, healthy), infectious (I, having the disease
and able to transmit it to others), or recovered 
(R, cured from the disease and immunized). This type of modeling gives a simplified description of the disease course, abstracting the continuous growth and decrease of the viral load and viral shedding of an individual. 
The modeling also defines the possible transitions between states: in the SIR model, an S individual can become I upon interaction with I individual(s), and an I individual becomes R upon recovery. In the SIS model instead, an I individual becomes again susceptible upon recovery. 

The second type of modeling hypothesis concerns the definition and representation of the interactions between individuals. This representation is crucial as it describes the way in which the process spreads between individuals. 
Numerous results have been obtained under the simplest homogeneous mixing hypothesis, in which any individual can interact with any other, and contagion occurs with a certain probability per unit time upon each contact \cite{anderson1992infectious,keeling2011modeling}. Even within this simplistic picture, the SIS and SIR models exhibit an interesting phenomenology, with 
a continuous phase transition at the so-called epidemic threshold:
when the ratio of the contagion to the recovery rate is smaller than the epidemic threshold, the spread dies out, while it
reaches a finite fraction of the population above the threshold. In the SIS case, a steady state is then reached, in which the epidemic is sustained by a non-zero number of individuals.

One of the most successful impacts of network science has been to go beyond the homogeneous mixing hypothesis and study how  more realistic structures of interactions between individuals affect the dynamics of compartmental models of contagion processes, and in particular 
the epidemic threshold \cite{pastor2001epidemic,barrat2008dynamical,pastor2015epidemic,kiss2017mathematics}. Indeed, network-based representations are conveniently used to describe many systems of various nature, including the social structures on which
many dynamical processes occur, such as the spread
of diseases and of information, the formation of opinions  and the diffusion of innovations \cite{albert2002statistical,newman2003structure,barrat2008dynamical}.
In the resulting modeling, the transmission process is assumed
to occur through pairwise interactions and through a single exposure: in other words, an infectious individual can transmit the disease to a susceptible one upon a single interaction
(along one of the links of the network representation). 

While such ``simple contagion'' frameworks are still widely used in the modeling of infectious diseases, the situation is more complex when dealing with social contagion phenomena, such as the adoption of norms or new products, or the diffusion of rumors. Indeed, empirical evidence has shown that simple
epidemic-like contagion processes do not provide a satisfactory description of the complex dynamics occurring when peer influence and reinforcement mechanisms are at work
\cite{centola2007complex,centola2010spread,ugander2012structural,weng2012competition,karsai2014complex,monsted2017evidence,guilbeault2018complex}. 
Complex contagion mechanisms have been proposed to account for these effects: broadly speaking, they are defined as any 
process in which exposure to multiple sources presenting the same stimulus is needed for the contagion to occur \cite{centola2007complex}. 
Modeling of complex contagion has been developed in two main directions. On the one hand, threshold models consider that 
an individual can be convinced to adopt e.g. a new behaviour if and only if a fraction of their contacts larger than a given threshold is already convinced (have already adopted the behaviour)
\cite{watts2002simple,centola2007complex,melnik2013multi,karsai2014complex,ruan2015kinetics,czaplicka2016competition}. On the other hand, epidemic-like processes have been generalized, with 
contagion rates that depend on the number of sources of exposure to which an individual is linked \cite{weng2012competition,cozzo2013contact,hodas2014simple,herrera2015understanding,o2015mathematical,czaplicka2016competition,tuzon2018continuous}.

From the homogeneous mixing simple contagion models to the complex contagion occurring on complex networks, the assumption of transmission processes occurring along pairwise interactions has remained an ubiquitous and most often undiscussed norm. It fits well with the representation of social groups as networks, since links of the networks are  pairwise  associations of nodes (the individuals of the population).
However, a number of social phenomena occur as the result of group interactions. Let us consider for instance the adoption of a product or a norm. An individual might be convinced by a single interaction with an adopter (simple contagion), or by successive interactions with two distinct adopters (complex contagion), along the links of their social networks. However, a qualitatively different process is at work if the individual gets convinced as part of a social group of three individuals, the other two being adopters. 
It might occur because the individual wants to be similar to the rest of the group, or, in a group discussion, the two adopters' arguments might reinforce each other in a way that would be impossible in separate pairwise discussions.

To account for such interactions between individuals occurring in groups of various sizes, it is thus necessary to expand the representation of the social structure from networks,
which can only encode pairwise interactions, 
to higher-order structures, namely hypergraphs \cite{battiston2020networks}: the building blocks of hypergraphs are indeed hyperedges that can join an arbitrary number of nodes. 
Clearly, the modeling of spreading processes on hypergraphs also implies to generalize contagion processes from pairwise to group processes: one needs for instance to define which contagion events can take place on a hyperedge joining $n$ nodes among which $m$ are infectious. 
A number of recent works have focused on the definition and
study of such models \cite{iacopini2019simplicial,de2020social,jhun2019simplicial,matamalas2020abrupt,de2020phase,landry2020effect}, and we review in this chapter some of the corresponding approaches and results, highlighting in particular how the obtained behaviour is richer than in the usual (network-based) contagion models.  The emerging phenomenology indeed includes both continuous and discontinuous transitions, hysteresis phenomena and critical mass phenomena reminiscent of the recently observed minimal size of committed minorities required to initiate social changes~\cite{centola2018experimental}.

\section{Spreading processes on higher-order structures}
\label{sec:model}

Group interactions can be encoded as hyperedges of an hypergraph, where each hyperedge is thus a set $[i_0, i_1,\dots,i_{k-1}]$ that involves $k$ elements. In this language, pairwise interactions are called $1$-hyperedges, 3-body interactions are called $2$-hyperedges, etc. 
In the broadest definition, there are no limitations to the size and relative inclusions of hyperedges. In some cases, it can be convenient to represent a social structure using the more restricted framework of simplicial complexes: such a representation assumes that in any group interaction all the sub-interactions among the group members should be considered as well~\cite{hatcher2002algebraic}. While this hypothesis has been used in Ref.~\cite{iacopini2019simplicial}, further developments have shown that similar dynamical outcomes for contagion processes can be found even under the more general framework of hypergraphs~\cite{de2020social}. Thus, in this chapter the latter setup will be used. 

As the interactions are not necessarily pairwise anymore, but can occur in groups of more than two individuals, this implies moreover that the models used to describe the contagion processes need to be redefined.
In this section, we present a rather general mathematical formulation
of such possible contagion models on higher-order
structures, defining it in terms of Bernoulli random variables and Poisson processes. 
Obtaining results directly from these definitions is, however, very hard, so that we mainly restrict this subsection to the definition of the models and of the quantities of interest, leaving to the following subsections the development of analytical approximations and the numerical simulations.

Mathematically, in the social contagion process the states of the nodes are modeled as Bernoulli random variables, $Y_i = 1$ (with its complementary $X_i = 0$) if the node is active and $Y_i = 0$ otherwise (and then  $X_i=1$). 
Individual states change either spontaneously or as a consequence of their interactions. Formally, this is a collection of independent Poisson processes. First,
we associate to
each active node $i$ a Poisson process with parameter $\delta_i$, modeling its spontaneous deactivation, $\{Y_i=1\} \xrightarrow[]{\delta_i} \{X_i=1\}$. 
This transition is similar to the healing in disease spreading dynamics. On the other hand, spreading processes occur along the hyperedges as follows. 
For each hyperedge $e_j$ we define a random variable $T_j = \sum_{k \in e_j} Y_k$: $T_j$ is by definition the number of active nodes in the hyperedge. If $T_j$ is equal to or above a given threshold $\Theta_j$, 
we model the contagion by a Poisson process with parameter $\lambda_j$. 
In other words, if $T_j \geq \Theta_j$, 
then $\{X_k=1\} \xrightarrow[]{\lambda_j} \{Y_k=1\} $, 
$\forall k \in e_j$.
This corresponds to a threshold process that becomes active only above a critical mass of active nodes. 
Finally, if $|e_j| = 2$, we assume directed Poisson processes, recovering a traditional SIS contagion process.
For the sake of simplicity, we assume that $\delta_i = \delta$ and $\lambda_j = \lambda \times \lambda^*(|e_j|)$, where $\lambda$ is the control parameter and $\lambda^*(|e_j|)$ is an arbitrary function of the cardinality of the hyperedge. The first assumption considers that every individual deactivates at the same rate. 
The second condition assumes that 
a hyperedge that is above its critical-mass threshold activates its nodes with a rate that
depends only of its cardinality (scaled by a global control parameter $\lambda$).
The exact equation describing the resulting dynamics can be written as
\begin{equation} \label{eq:exact}
 \dfrac{d \En{Y_i}}{dt}= 
 \En{-\delta Y_i + \lambda \left( 1 - Y_i\right) \sum_{e_j | i \in e_j} \lambda^*(|e_j|) \sum_{B} \mathbbm{1}_{\{ Y_i = 0, T_j \geq \Theta_j\}}},
\end{equation}
where the first summation is over all hyperedges containing node $i$ and the second over the 
set $B$ of all
possible dynamical micro-states inside the hyperedge $e_j$. Furthermore, $\mathbbm{1}_{\{ Y_i = 0, T_j \geq \Theta_j\}}$ is an indicator function depending on both the specific node and the hyperedge, taking the value $1$ if $Y_i = 0$ and the critical mass in the hyperedge is reached (i.e., if node $i$ is inactive and can potentially become active), and $0$ otherwise. We also use for convenience a global threshold ratio $\Theta^*$, with
 $\Theta_j = \ceil[\big]{\Theta^* |e_j|}$.

\begin{figure*}[t]
\includegraphics[width=\linewidth]{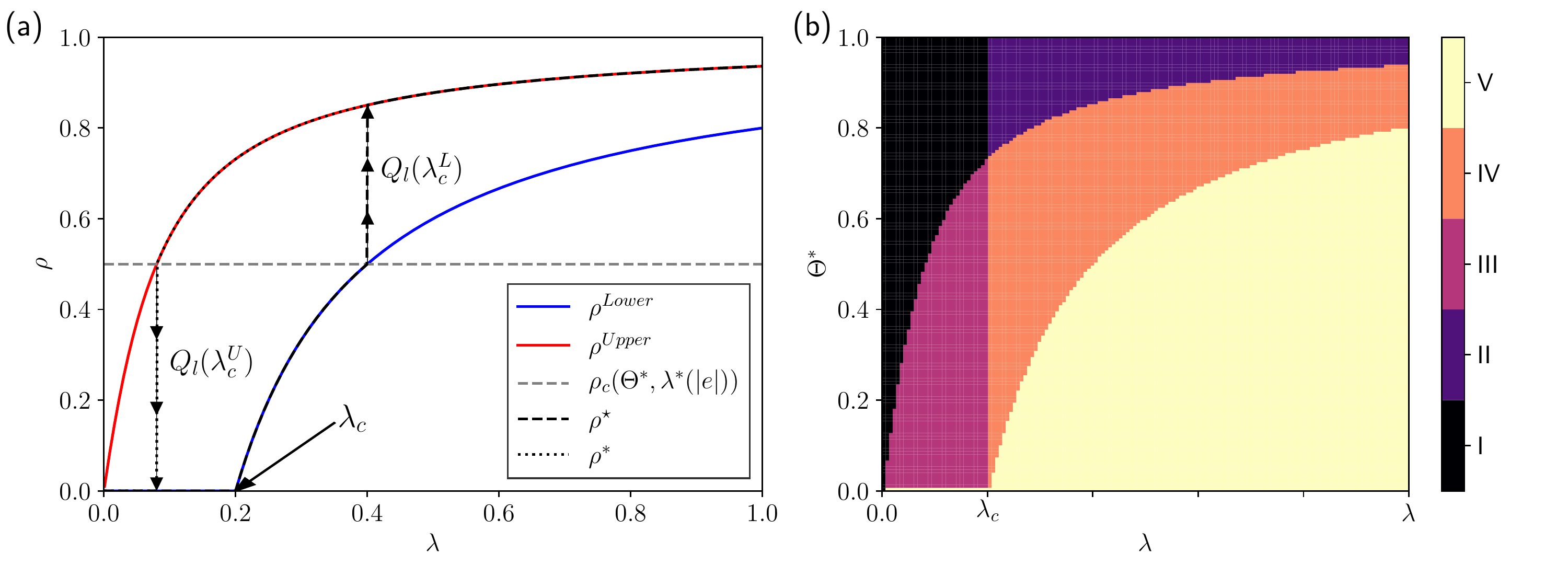}
\caption{Example of a phase diagram and parameter space for the hyperblob (See Section~\ref{sec:qmf_hyperblob} for details). 
Panel (a) shows the solutions for a fixed $\Theta^*=0.5$. 
The red and blue curves show respectively 
$\rho^{\text{Upper}}$ and $\rho^{\text{Lower}}$.
The lower solution presents a second-order phase transition at $\lambda_c = 0.2$.
When increasing $\lambda$ from $0$ to $1$
the transition from the lower to the upper solution 
 occurs at the intersection of the lower solution with a value of $\rho$ in which the upper solution becomes the only stable one, $\rho_c$. The jump between the two solutions is  $Q_l(\lambda_c^{L})$.
 Similarly, when decreasing back $\lambda$, the jump from
 $\rho^{\text{Upper}}$ to $\rho^{\text{Lower}}$ takes
 place when $\rho^{\text{Upper}}$ crosses the value 
 $\rho_c$ and becomes unstable, and the density jump is
 $Q_l(\lambda_c^{U})$. 
 Panel (b) shows a sketch of the parameter space. Region I: the system reaches the absorbing state, $\rho=0$; Region II:  only the lower solution is stable; Region III: $\rho^{Upper}$ is stable and $\rho^{\text{Lower}} = 0$ (bistable region below the critical point); Region IV: $\rho^{Upper} > \rho^{Lower} > 0$ and both are stable (bi-stable); Region V: only the upper solution is stable.}
\label{fig:schematic}
\end{figure*}

The order parameter is defined as the expected fraction of active nodes, i.e., $\rho = \frac{1}{N} \sum_i \En{Y_i}$. Although a formal proof is yet lacking, we observed, through simulations and numerical solutions of several analytical approaches, a rather general phenomenology when $\lambda$ is varied at fixed $\Theta^*$, as illustrated
in Figure \ref{fig:schematic}. Two solutions for $\rho$ as a function of $\lambda$ are generically obtained, here called $\rho^{\text{Lower}}$ and $\rho^{\text{Upper}}$
($\rho^{\text{Upper}} > \rho^{\text{Lower}}$). 
Moreover, under certain conditions, the $\rho^{\text{Lower}}$ solution presents a continuous phase transition between the absorbing state, where all the individuals are deactivated
($\rho^{\text{Lower}}=0$), and an active state
($\rho^{\text{Lower}}>0$). This transition occurs
at a critical value of the parameter $\lambda$ 
denoted $\lambda_c$. 
A bistable region can also exist, in which the
final state depends on the initial condition $\rho(t=0)$ 
being below or above a so-called global critical-mass denoted $\rho_c$. 
Let us denote by $\rho^{\bigstar}$ 
the solution that is obtained if $\rho(t=0) < \rho_c$ and by $\rho^{*}$ solution 
obtained if $\rho(t=0) \geq \rho_c$. In the bistable region
$\lambda_c^{\text{U}} < \lambda < \lambda_c^{\text{L}}$,
$\rho^{\bigstar}= \rho^{\text{Upper}}$
and $\rho^{*} = \rho^{\text{Lower}}$, while 
for $\lambda < \lambda_c^{\text{U}}$
we have $\rho^{\bigstar}= \rho^{*} = \rho^{\text{Lower}}$, 
and for $\lambda >  \lambda_c^{\text{L}}$,
$\rho^{\bigstar}= \rho^{*} = \rho^{\text{Upper}}$.

When increasing $\lambda$ from $0$ (forward phase diagram), the system thus first follows $\rho^{\text{Lower}}$ and jumps to 
$\rho^{\text{Upper}}$ at $\lambda_c^{\text{L}}$
when $\rho^{\text{Lower}}$ becomes unstable. When
decreasing back $\lambda$ (backward phase diagram), 
the system follows $\rho^{\text{Upper}}$ and jumps
back to $\rho^{\text{Lower}}$  at $\lambda_c^{\text{U}}$
where $\rho^{\text{Upper}}$ becomes unstable.
The length of these two jumps are defined 
 as
\begin{equation}
 Q_l(\lambda_c^{X}) = \left( \rho^{\text{Upper}} - \rho^{\text{Lower}} \right)_{\lambda = \lambda_c^X},
\end{equation}
where $Q_l(\lambda_c^{X})$ can be $Q_l(\lambda_c^{\text{L}})$ or $Q_l(\lambda_c^{\text{U}})$. These quantities give
the sudden change in the fraction of active nodes at these
jumps. 
These concepts are exemplified in Fig.~\ref{fig:schematic}(a), where we show an example obtained for a homogeneous hypergraph composed of a random regular network and a hyperedge containing all the nodes. This structure's symmetries allow us to analytically explore their solutions following a first-order approximation and serve as a didactic example of the behaviors present in our model. In Fig.~\ref{fig:schematic}(b), we show a sketch of the
$(\lambda, \Theta^*)$ parameter space for the same structure.
We present the analytical aspects of this solution in Section~\ref{sec:qmf_hyperblob}.

\section{Individual-based or quenched mean-field approach}

As mentioned above, the exact formulation provides only a conceptual understanding of our model but fails to provide a quantitative characterization. Here we consider the individual-based or also called quenched mean-field approximation. This approach neglects dynamical correlations but takes into account the structural correlations of the interactions of the nodes. It is possible to solve the resulting equations numerically (without resorting to stochastic numerical simulations), obtaining a better understanding of the model's behaviour. We first derive the general dynamical equations of this approximation in Section~\ref{sec:qmf_general}; we then consider a toy example and solve numerically the corresponding equations in Section~\ref{sec:qmf_hyperblob} in order to exemplify the variety of behaviors present in our model. Finally, in Section~\ref{sec:qmf_pl} we consider a hypergraph with power-law distributed cardinalities of hyperedges, which has a  more complex and heterogeneous structure than the toy example of Section~\ref{sec:qmf_hyperblob}.

\subsection{The general formulation}
\label{sec:qmf_general}

Since Eq.\eqref{eq:exact} cannot be numerically solved, here we assume that the random variables are independent, allowing us to significantly reduce the complexity of our model.  Denoting $y_i = \En{Y_i}$, this first-order approximation is given by
\begin{equation} \label{eq:first_order}
  \dfrac{d y_i}{dt}= -\delta y_i + \lambda \left( 1 - y_i\right) \sum_{e_j | i\in e_j } \sum_{k = \Theta_j}^{|e_j|} \lambda^* (|e_j|) \mathbb{P}_{e_j} \left(K=k \right),
\end{equation}
where $\mathbb{P}_{e_j} \left(K=k \right)$ is the probability that the hyperedge $e_j$ has $k$ active nodes. In this formulation, we have used that the expectation of the indicator function in Eq.~\eqref{eq:exact} follows a Poisson binomial distribution, which can be formally expressed as
\begin{eqnarray}
 \En{\mathbbm{1}_{\{ (T_j - Y_k) \geq \Theta_j\}}} \approx \sum_{m = \Theta_j}^{|e_j|} \mathbb{P}_{e_j} \left(K=m \right) \\
 \mathbb{P}_{e_j} \left(K=m \right) = \sum\limits_{A\in F_m} \prod\limits_{i\in A} y_i \prod\limits_{i'\in A^c} (1-y_{i'}), \label{eq:prob_pb}
\end{eqnarray}
where $F_m$ is the set of all subsets of $m$ integers in $\{1, 2, ... ,|e_j|\}$ 
and $A^c$ is the complementary of $A$.
The summation in  Eq.~\eqref{eq:prob_pb}
considers all possible micro-configurations in a given hyperedge, with 
$A$ accounting for the active nodes and $A^c$ for the  inactive ones. Using it directly for numerical computations
can introduce numerical stability problems for large
hyperedges~\cite{fernandez2010closed}.  Fortunately, this issue can be solved by considering the discrete Fourier transform, obtaining the following numerically stable solution~\cite{fernandez2010closed}:
\begin{equation} \label{eq:dft_pn}
 \mathbb{P}_{e_j} \left(K=k \right) = \frac{1}{n+1} \sum\limits_{l=0}^n C^{-lk} \prod\limits_{m=1}^n \left( 1+(C^l-1) y_m \right),
\end{equation}
where $C=\exp \left( \frac{2i\pi }{n+1} \right)$.
This expression allows to compute the solution for arbitrarily large hyperedges. Although the whole argument is quite intricate, Eq.~\eqref{eq:dft_pn} is simple and robust enough, allowing the numerical evaluation of Eq.~\eqref{eq:first_order} for arbitrary hypergraphs and parameters.

\subsection{The hyperblob}
\label{sec:qmf_hyperblob}

For the sake of simplicity, let us focus here on a very particular and homogeneous structure: the hyperblob. The hyperblob is a hypergraph constructed as a homogeneous set of pairwise interactions with average degree $\langle k \rangle$, to which a single additional hyperedge containing all nodes is added. This structural simplicity 
allows us to solve the model analytically. Indeed,
given the symmetry of the system, all $y_i$ are equal
($y_i=\rho \ \forall i$ )
and their evolution can be expressed by the
following single equation:
\begin{equation} \label{eq:rrn}
 \dfrac{d \rho}{dt} = -\delta \rho + \lambda (1 - \rho) \left[ \langle k \rangle \rho + \lambda^* F \left( \Theta^*, \rho \right) \right].
\end{equation}
Here $\lambda^*$ stands for $\lambda^*(|e_j|)$ and
\begin{equation} \label{eq:F}
 F \left( \Theta^*, \rho \right) = 1 - \sum_{l=0}^{\Theta-1} \mathbb{P}_{N-1} \left(K=l \right) \approx
 \begin{cases}
  1, \hspace*{0.3cm} \text{if} \hspace*{2mm} \rho \geq \Theta^* \\
  0, \hspace*{0.3cm} \text{otherwise}
 \end{cases},
\end{equation}
where the approximation on the right-most part of the equation assumes that the hypergraph is sufficiently large (for more on this approximation, we refer to the supplemental material of~\cite{de2020social}). 

The approximation in Eq.~\eqref{eq:F} suggests
the possibility of having two solutions, one such that $F \left( \Theta^*, \rho \right) = 0$, and another one such that $F \left( \Theta^*, \rho \right) = 1$. Note that the first solution represents the case in which the largest hyperedge is inactive, while it is active in the second case. We remark that, as the only higher-order structure contains all the nodes, the global critical-mass is here 
$\rho_c = \Theta^*$. In other words, the activation of this hyperedge determines which solution the system is in, and
the jumps between upper and lower solutions happen when they ``cross'' the value $\rho = \Theta^*$. 
From the approximation in Eq.~\eqref{eq:F} and Eq.~\eqref{eq:rrn}, we can analytically obtain
the model's parameter space, obtaining the two solutions
(see \cite{de2020social})
\begin{eqnarray}
& &\rho^{\text{Lower}} = 
 \begin{cases}
  1 - \frac{\delta}{\langle k \rangle \lambda}, \hspace*{1cm} \text{if} \hspace*{2mm} \frac{\lambda}{\delta} \geq \frac{1}{\langle k \rangle}\\
  0, \hspace*{2.1cm} \text{otherwise}
 \end{cases}\\
  & &\rho^{\text{Upper}} = \frac{-\delta + \langle k \rangle \lambda - \lambda^* \lambda + \sqrt{ 4 \langle k \rangle \lambda^* \lambda^2 + (\delta + (-\langle k \rangle + \lambda^*) \lambda)^2}}{(2 \langle k \rangle \lambda)}.
\end{eqnarray}
As anticipated in Section \ref{sec:model}, 
a second-order phase transition is obtained 
for $\rho^{\text{Lower}}$
as the feasibility condition $\frac{\lambda}{\delta} \geq \frac{1}{\langle k \rangle}$~\cite{de2018fundamentals}.
We remark that this lower solution is here simply the solution of a mean-field approach for an homogeneous structure with average degree $\langle k \rangle$. The next quantity of interest are the limits of the bistable region, 
which can be calculated as 
\begin{eqnarray}
 \lambda_c^{\text{L}} &=& \frac{\delta}{\langle k \rangle - \Theta^*  \langle k \rangle} \\
 \lambda_c^{\text{U}} &=& -\frac{\delta  \Theta^* }{\lambda^*  \Theta^* -\lambda^* +(\Theta^*)^2 \langle k \rangle - \Theta^* \langle k \rangle}.
\end{eqnarray}
Finally, the jump length is expressed as
\begin{equation}
 Q_l(\lambda_c^{X}) = \left(\frac{\delta -\lambda  (\lambda^* +\langle k \rangle) +\sqrt{(\delta +\lambda  (\lambda^* -\langle k \rangle))^2+4 \lambda^*  \langle k \rangle \lambda^2}}{2 \langle k \rangle \lambda } \right)_{\lambda = \lambda_c^X},
\end{equation}
where $\lambda_c^{X}$ can be $\lambda_c^{\text{L}}$ or $\lambda_c^{\text{U}}$. 
Although these equations are reasonably simple, the upper solution depends on a quadratic equation, where only one of the solutions is physical. The details for the complete derivation of these results can be found in the supplemental material of~\cite{de2020social}. In the same reference, the interested reader can also find similar results when 
the considered lower-order structure is a star graph.

\begin{figure*}[t]
\includegraphics[width=\linewidth]{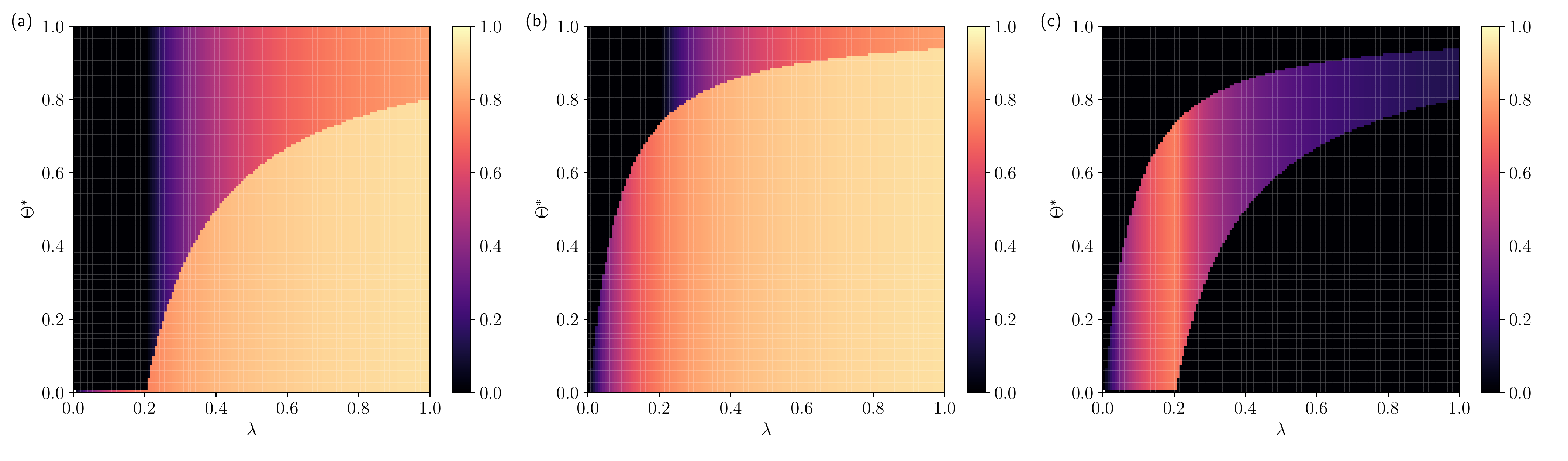}
\caption{Phase diagram for the Hyperblob with $\langle k \rangle = 5$, $\delta = 1$ and $\lambda^*(|e_j|) = \log_2(|e_j|)$. In (a)-(c) the colormaps are obtained changing $\lambda$ and $\Theta^*$. In (a) the solution of forward phase diagram, in (b) the solution of the backward one and in (c) the jump length (i.e., difference between (b) and (a)), emphasizing the bi-stability region.}
\label{fig:analytical_pd_rrn}
\end{figure*}

Figure~\ref{fig:analytical_pd_rrn} shows the phase diagram for a hyperblob with $N = 10^3$, $\langle k \rangle = 5$, $\delta = 1$ and $\lambda^*(|e_j|) = \log_2(|e_j|)$. This result complements Fig.~\ref{fig:schematic} and exemplifies the five regions of the diagram in Fig.~\ref{fig:schematic}(b). As predicted by our solutions, in (a), we observe that the absorbing state plays a major role in the forward diagram, $\rho^{\bigstar}$, as there is a region of the parameter space that is not active, which is a consequence of the second-order phase transition present in $\rho^{\text{Lower}}$. 
This is also depicted as Regions I and III in Fig.~\ref{fig:schematic}(b). Conversely, for the backward phase diagram, $\rho^{*}$, the set of parameters in which the system can reach the absorbing state is rather reduced, being restricted to Region I in Fig.~\ref{fig:schematic}(b). We highlight that substituting the random regular network by a star would slightly change the parameter space as the second-order phase transition of $\rho^{\text{Lower}}$ vanishes in the limit $N \rightarrow \infty$, thus implying that the Regions I and III in Fig.~\ref{fig:schematic}(b)  vanish as well.

\subsection{Example of a hypergraph with a power-law distribution of cardinalities}
\label{sec:qmf_pl}

\begin{figure}
	\centering
	\includegraphics[width=\textwidth]{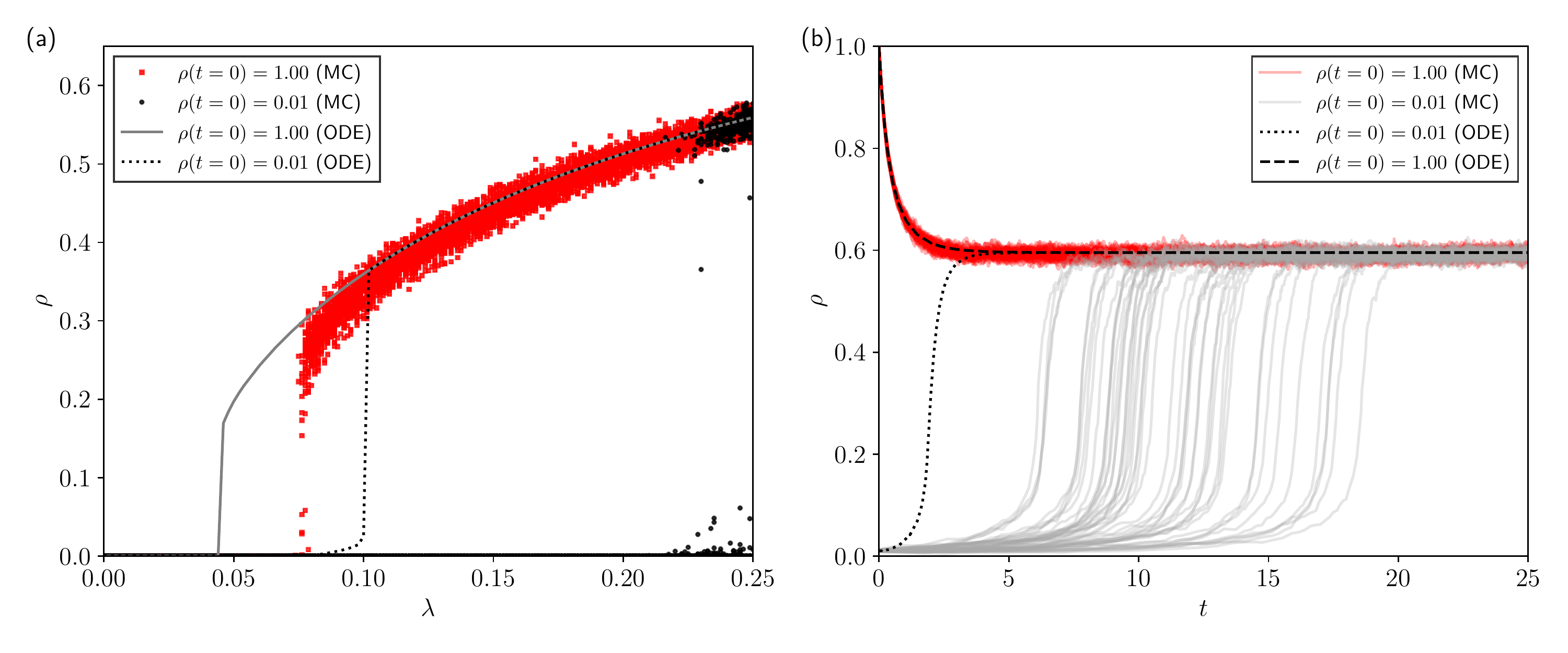}
	\caption{Comparison of the numerical solution of the ODE system Eq.~\eqref{eq:first_order} with Monte Carlo (MC) simulations for a hypergraph with $N = 10^4$, power-law distributed cardinalities, $P(|e_j|) \sim |e_j|^{-\gamma}$ with $\gamma = 2.25$, and $\min \{ |e_j| \} = 2$. The spreading rates are taken as $\lambda_j = \lambda \times \log_2(| e_j |)$ and the deactivation parameter is fixed as $\delta = 1$. We show in panel (a) shows the phase diagram for both the upper and lower solutions, and in panel (b) the temporal behavior for $\lambda = 0.25$.}
	\label{fig:MC_ODE_PL_Ex}
\end{figure}

In order to consider more complex and heterogeneous structures, we show in Figure~\ref{fig:MC_ODE_PL_Ex} an example of the solutions of the system of equations~\eqref{eq:first_order} for a hypergraph with $N = 10^4$, power-law distributed cardinalities, $P(|e_j|) \sim |e_j|^{-\gamma}$ with $\gamma = 2.25$, and $\min \{ |e_j| \} = 2$. 
Here we use spreading rates $\lambda_j = \lambda \times \log_2(| e_j |)$ and, we fix the deactivation parameter as $\delta = 1$. 
Figure~\ref{fig:MC_ODE_PL_Ex}(a) shows the phase diagram, while Figure~\ref{fig:MC_ODE_PL_Ex}(b) displays the temporal behavior for $\lambda = 0.25$. The agreement is qualitatively good, with the upper solution being well captured.
Both $\lambda_c^{\text{U}}$ and
$\lambda_c^{\text{L}}$ seem to be underestimated by the ODE. 
We however remark that an accurate determination of the transition points from numerical simulations is not an easy task and requires
 more sophisticated algorithms (see Ref.~\cite{de2020social} and its supplemental material for more details). 
 The temporal behavior shown in Fig.~\ref{fig:MC_ODE_PL_Ex}(b) suggests that the upper solution is better captured by our approach even at the dynamical level. For the lower solution, the steady-state value is very well captured, but the duration of the transient is longer in Monte Carlo simulations.

In summary, these examples highlight 
that the first-order approximation can provide
a qualitative picture of the phenomenology at work, but that
its limitations still need to be further evaluated. 
For instance, the accuracy of the estimated discontinuities' might be related to specificities of the considered structure (e.g., low average degree or hyperedge intersections). 
The strong interest of this approximation lies in the
relatively easy numerical implementation, as the system
 of equations~\eqref{eq:first_order} can be solved, e.g.,
 by using Runge-Kutta methods. The qualitative picture obtained also suggests that the first-order approximation might be a good starting point for further analytical explorations of this type of models.

\section{Annealed mean-field approach}

\subsection{Homogeneous mean-field}
We now focus on the simplest analytical framework, the mean-field (MF) approach, in which we assume that the population is fully mixed such that nodes are statistically equivalent, their states are independent, and all interactions can happen with identical probabilities. This is indeed the simplest scenario, which completely neglects the underlying structure. 
The mean-field form of Eq.~\eqref{eq:exact} is given by
\begin{equation} \label{eq:poly}
    \dfrac{d \rho}{dt} = -\delta \rho + P^\nu(\rho) = -\delta \rho + \sum_{m=2}^\nu \lambda_m c(m) (1 - \rho) \rho^{m-1},
\end{equation}
where $\nu = \max_j \{| e_j| \}$ is the maximum cardinality and $c(m)$ is the ratio between the average number of hyperedges with cardinality $m$ and the average number of pairwise interactions incident on a node $i$, which characterizes the structure of the hypergraph. In the steady-state this is a polynomial equation whose solutions are the fixed points of the process.

We now restrict our attention to a tractable case in which we can find a solution to the MF approximation. To do this, we consider a hypergraph formed solely by 1-hyperedges (standard pairwise links) and 2-hyperedges (3-body interactions). In this case, the maximum cardinality is $\nu = 3$, and Eq.~\eqref{eq:poly} simplifies to:
\begin{equation} \label{eq:poly:nu3}
    \dfrac{d \rho}{dt} = -\delta \rho +\lambda_2 c(2) (1 - \rho)\rho+\lambda_3 c(3) (1 - \rho)\rho^2.
\end{equation}
Notice that $c(2)=1$ by definition, while $c(3)$ is given by the ratio between the average number of 2-hyperedges and the average number of 1-hyperedges adjacent to a node, so that $c(3) = \langle k_3\rangle/\langle k_2\rangle$.
We can thus rewrite Eq.~\eqref{eq:poly:nu3} as
\begin{equation} \label{eq:MF:lambdas}
\dfrac{d \rho}{dt} = -\delta \rho + \lambda_2(1 - \rho)\rho+ \frac{\langle k_3\rangle}{\langle k_2\rangle}\lambda_3 (1 - \rho)\rho^2.
\end{equation}
After defining $\beta_2 = \lambda_2\langle k_2\rangle/\delta$ and $\beta_3 = \lambda_3\langle k_3\rangle/\delta$, we can rewrite Eq.~\eqref{eq:MF:lambdas} as:
\begin{equation}\label{eq:MF:betas}
\dfrac{d \rho}{dt} = -\rho + \beta_2(1 - \rho)\rho+ \beta_3 (1 - \rho)\rho^2.
\end{equation}

From Eq.~\eqref{eq:MF:betas} it is evident that we can recover the standard MF equation for the SIS model by setting $\beta_3=0$. In this case, we get back the two standard stationary solutions which correspond to the absorbing state with no infected nodes $\rho^{* \left[\beta_3=0\right]}_1=0$ and the endemic state $\rho^{* \left[\beta_3=0\right]}_2 = 1 - 1/\beta_2$. When $\beta_2<1$, $\rho^{* \left[\beta_3=0\right]}_1$ is the only (stable) solution;
it becomes unstable when $\beta_2>1$ and $\rho^{* \left[\beta_3=0\right]}_2$ appears (stable). The standard epidemic threshold $\beta_2=1$ represents the points at which the system undergoes a continuous transition between the two regimes.

Let us now consider the more interesting case in which there are contributions coming from the higher-order interactions (2-hyperedges), i.e., $\beta_3>0$. In this case, 
there are up to three stationary solutions of
the steady state equation $d_t \rho=0$
that fall within the range $\rho\in[0,1]$. One is the trivial solution $\rho^{*}_1=0$, which corresponds to the usual absorbing state where the epidemics dies out. The other two non-trivial solutions are given by
\begin{equation}\label{eq:MF:sol2}
\rho^{*}_{2\pm} = \frac{ \beta_{3}-\beta_2 \pm \sqrt{(\beta_2 - \beta_{3})^2 -4\beta_{3}(1-\beta_2)} }{2\beta_{3}} .
\end{equation}

These correspond to the lower ($\rho^{*}_{2-}$) and upper ($\rho^{*}_{2+}$) branch that have been previously discussed. This simple mean-field description allows to go further and study the stability of the system (see \cite{iacopini2019simplicial} for details), confirming that:

\begin{itemize}
    \item When $\beta_3\leq 1$, if $\beta_2<1$ there is only one acceptable solution, that is the trivial absorbing state $\rho^*_1=0$. If instead $\beta_2>1$, the non-trivial solution $\rho^{*}_{2+}$ is positive and stable, while $\rho^*_1$ becomes unstable. Thus, when moving--using the standard control parameter--from $\beta_2<1$ to $\beta_2>1$, it is possible to show that the system undergoes a continuous transition at the standard epidemic threshold $\beta_2=1$. While this is similar to what happens when $\beta_3=0$ (standard SIS model), if $0<\beta_3\leq 1$ there is a higher density of infected nodes in the endemic state.\\
    
    \item When $\beta_3> 1$, algebraic manipulations of Eq.~\eqref{eq:MF:sol2} shows that if $\beta_2<\beta_c = 2\sqrt{\beta_3}-\beta_3$, $\rho^{*}_{2\pm}$ are not in the acceptable domain and the only (stable) solution is, again, the trivial one $\rho^*_1=0$. Contrarily, if $\beta_2 >\beta_c$, the system presents two different regimes. If $\beta_2>1$, we have a scenario similar to the one above, where $\rho^*_1$ is unstable and the stable state is $\rho^{*}_{2+}>0$. If instead $\beta_c<\beta_2 <1$, both solutions $\rho^{*}_{2\pm}$ are positive ($0 < \rho^{*}_{2-}<\rho^{*}_{2+}$). More precisely, $\rho^{*}_{2-}$ is an unstable solution that splits the phase space into two regions and determines--according to the initial conditions--in which one of the other two stable solutions $\rho^{*}_1$ and $\rho^{*}_{2+}$ the system will end up. We can thus confirm what we had previously observed, that is the presence of a discontinuous transition at $\beta_c$
    and of a bistable region in which the system reaches
    $\rho^{*}_{2+}$ only
    if the initial seed of infected nodes is above a critical mass ($\rho(t=0)>\rho^{*}_{2-}$).
\end{itemize}

\begin{figure}[t]
	\centering
	\includegraphics[width=0.55\textwidth]{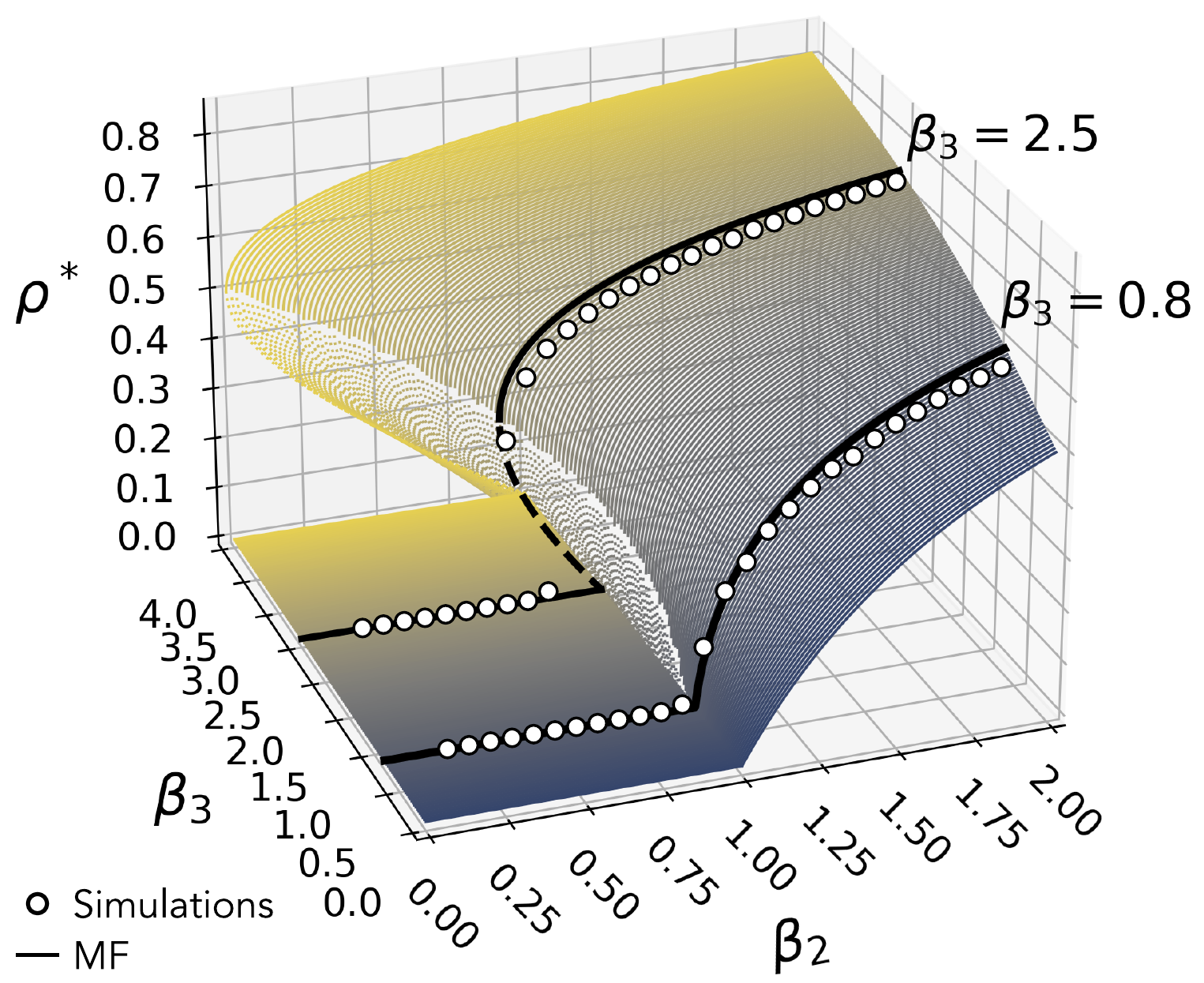}
	\caption{Analytical solution in the mean-field approximation.
	Three-dimensional phase diagram, where the density of infected nodes in the stationary state $\rho^*$ is plotted as a function of the 1-hyperedges rescaled infectivity $\beta_2=\lambda_2\langle k_2\rangle/\delta$ and the 2-hyperedges rescaled infectivity $\beta_3 = \lambda_3\langle k_3\rangle/\delta$. When $\beta_3=0$ the dynamics obeys the one of the standard SIS model on networked systems with no higher-order interactions (links only). Two example curves are shown (at constant values of $\beta_3=0.8$ and $\beta_3=2.5$), where MF results (black lines) are compared to results of stochastic simulations on random simplicial complexes (white circles)~\cite{iacopini2019simplicial}.}
	\label{fig:MF_3D}
\end{figure}

These results are also illustrated in Fig.~\ref{fig:MF_3D}, which gives a three-dimensional representation of the phase diagram associated to the system. These are the solutions of Eq.~\eqref{eq:MF:betas} just described, representing the density of infected nodes in the large time limit as a function of the rescaled infectivity parameters $\beta_2$ and $\beta_3$. For visualization purposes only stable solutions are shown when $\beta_2>1$. We also plot two representative curves (black lines) that highlight the possible types of transitions. For $\beta_3=0.8$ the system still presents the standard continuous transition at $\beta_2=1$, while for greater values ($\beta_3=2.5$ shown in the figure) the transition becomes discontinuous. The presence of a bistable region is evident from the ``folding'' of the surface, in which a line parallel to the vertical axes can cross the surface in two distinct points.

We also compare the MF results with average stationary values extracted from multiple runs of stochastic simulations (white circles). Notice how, despite the oversimplified MF approach, the analytical predictions--on the position of the epidemic threshold and the nature of the transition--are in good agreement with the simulations when higher-order structures with homogeneous degree distributions are considered, such as the random simplicial complex structure used in this case ($N=2000$, $\langle k_2\rangle=20$, $\langle k_3\rangle=6$). More details on the construction of this random structure are given in Ref.~\cite{iacopini2019simplicial}. 

\medspace 

\subsection{Heterogeneous mean-field}

The MF approach can be improved by relaxing the assumption that all nodes are equivalent, and considering instead that nodes within the same hyperdegree class behave similarly~\cite{pastor2001epidemic}. Let us thus call $\vec{k}_i$ the vector containing all the generalized degrees associated to node $i$ up to the maximum cardinality $\nu$, such that $\vec{k}_i=[k_{2,i},k_{3,i},\dots,k_{\nu,i}]$.
By doing that, we are effectively removing the actual structure and describing it in terms of the probabilities of nodes to share a hyperedge.
The equation for the heterogeneous mean-field (HMF) approach, as introduced in Ref.~\cite{landry2020effect}, reads:

\begin{eqnarray}\nonumber
\dfrac{d \rho_{\vec{k}}}{dt} &=& -\delta \rho_{\vec{k}} + 
(1-\rho_{\vec{k}}) \times \\
\sum^{\nu}_{m=2}&\frac{\lambda_m}{(m-1)!}&
\sum_{\vec{k}_1,\dots,\vec{k}_m}\prod^{m-1}_{\ell}P(\vec{k}_\ell)f_m(\vec{k},\vec{k}_1,\dots,\vec{k}_{m-1})G(\rho_{\vec{k}_1},\dots,\rho_{\vec{k}_{m-1}})
\label{eq:HMF}
\end{eqnarray}
where $\rho_{\vec{k}}$ denotes the density of active nodes having hyperdegree $\vec{k}$, and $P(\vec{k})$ the number of nodes with hyperdegree $\vec{k}$. In the second term of the r.h.s. of Eq.~\eqref{eq:HMF}, the first summation runs over all hyperedges of size $m$ that can infect a node having hyperdegree $\vec{k}$. This means that for each hyperedge there are $m-1$ other nodes that could be infected, and their combinations are counted by the second summation. The ability to actually transmit the infection depends on the fraction of hyperedges (among all their possible combinations) that include the given node, given by $f_m(\vec{k},\vec{k}_1,\dots,\vec{k}_{m-1})$, and the probability $G(\rho_{\vec{k}_1},\dots,\rho_{\vec{k}_{m-1}})$ that the given hyperedge can transmit the infection. If we assume that a hyperedge can infect a node only if all the remaining nodes composing it are infected, this reads $G(\rho_{\vec{k}_1},\dots,\rho_{\vec{k}_{m-1}})=\prod^{m-1}_{\ell=1}\rho_{\vec{k}_\ell}$. In addition, if we consider as before a hypergraph containing 1- and 2-hyperedges only ($\nu=3$), and we assume that the connection probabilities are only determined by the links, i. e., $f_m(\vec{k},\vec{k}_1,\dots,\vec{k}_{m-1})=f_m(k,k_1,\dots,k_{m-1})$, Eq.~\eqref{eq:HMF} simplifies to

\begin{equation}\label{eq:HMF:simple}
\begin{split}
\dfrac{d \rho_k}{dt} &= -\delta \rho_k + (1-\rho_k)\lambda_2\sum_{k_1}P(k_1)f_2(k,k_1)\rho_{k_1}\\
&+(1-\rho_k)\frac{\lambda_3}{2}\sum_{k_1,k_2}P(k_1)P(k_2)f_3(k,k_1,k_2)\rho_{k_1}\rho_{k_2}
\end{split}
\end{equation}
where it is now possible to explicitly distinguish the contributions coming from links and ``triangles'', respectively the second and third term of the r.h.s. of Eq.~\eqref{eq:HMF:simple}.

The process described by Eq.~\eqref{eq:HMF:simple} can be analyzed using linear stability analysis. Although an analytical solution for the fixed points of Eq.~\eqref{eq:HMF:simple} is not possible, we can restrict our analysis to the inactive state, i.e., $\rho_k = 0$ for all $k$. And as it turns out~\cite{landry2020effect}, the inactive state becomes unstable for
\begin{equation}
\frac{\lambda_2}{\delta} > \frac{\langle k_2\rangle}{\langle k_2^2\rangle},
\end{equation}
where $k_2$ is the pairwise degree.
Interestingly, the take-home message from this analysis is that only pairwise interactions are responsible for the inactive state's stability. In this case, the parameter $\lambda_3$ is responsible for the presence or absence of bi-stable solutions. As $\lambda_3$ increases, the dynamics allow for a discontinuity, bi-stability, and hysteretic behavior. These results are in agreement with the approach of Ref.~\cite{de2020phase}, where the authors arrived at a similar conclusion using a quenched formalism. Furthermore, in Ref.~\cite{landry2020effect}, the authors used the HMF formalism to investigate the effect of heterogeneity in hypergraph contagion models. They showed that in the extreme case where a hyperedge can transmit infection if there is at least one infectious node (as opposed to $m-1$ discussed here), the bi-stability disappears, and the critical point depends on both $\lambda_2$ and $\lambda_3$. Interestingly, they also showed that the explosive transition could disappear for specific heterogeneous structures, e.g., when power-law distributions of pairwise interactions are used as a starting structure to construct the hypergraph. This can also happen when 2-hyperedges are placed at random, as opposed to degree-correlated structures where higher-order interactions are more likely to involve nodes that have a high pairwise degree (more details on the effects of heterogeneity and the HMF formalism can be found in Ref.~\cite{landry2020effect}).

\section{Simulations on real-world structures}
\label{sec:Simulations}

\begin{figure}[]
	\centering
	\includegraphics[width=\textwidth]{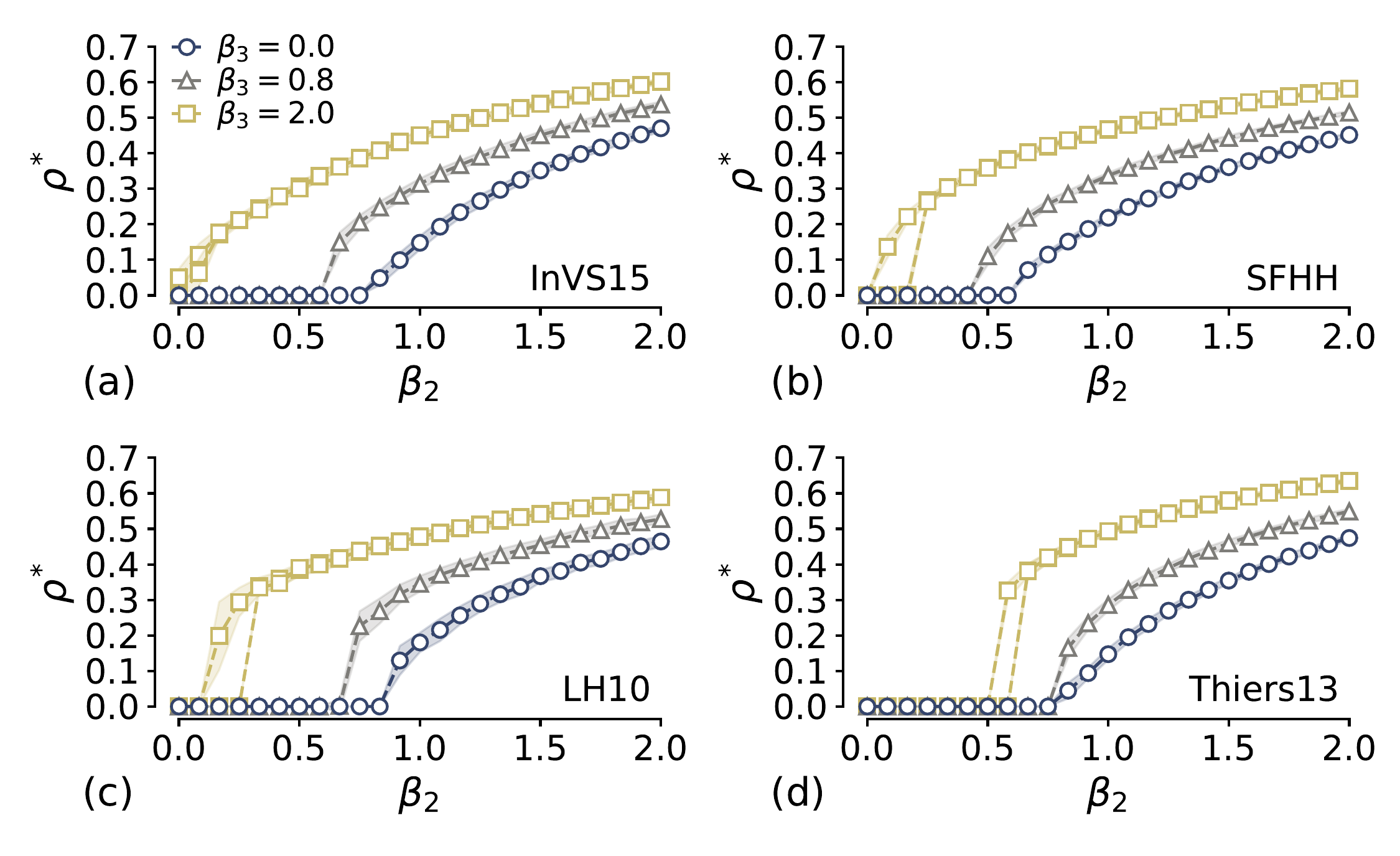}
	\caption{Simulations on real-world social structures. Density of infected nodes $\rho^*$ in the stationary state as a function of the 1-hyperedges rescaled infectivity $\beta_2=\lambda_2\langle k_2\rangle/\delta$ and for three different values of the 2-hyperedges rescaled infectivity $\beta_3 = \lambda_3\langle k_3\rangle/\delta$. When $\beta_3=0$ the dynamics obeys the one of the standard SIS model on networked systems with no higher-order interactions. Points and shaded areas correspond to median values and standard deviations as extracted from stochastic simulations on top of four different empirical simplicial complexes constructed from the SocioPatterns data sets: a workplace (a), a conference (b), a hospital (c) and a high school (d).
		See \cite{iacopini2019simplicial} for details.
	}
	\label{fig:sociopatterns}
\end{figure}

While the analytical approximations developed in the above sections correspond to simplified structures of interactions between nodes, real-world interactions are expected to involve complex and intricate structural correlations at various scales that are not easily reproduced by models. Therefore, we now briefly investigate the dynamics of the higher-order social contagion model on empirical higher-order structure. We focus in particular on the simplicial contagion model in its original formulation, where the social structure is modeled as a simplicial complex and each simplex of size $k$ can transmit the infection (at its order-dependent rate) to a susceptible node incident on it only if the remaining $k-1$ nodes are infectious~\cite{iacopini2019simplicial}.

To this aim, we construct empirical simplicial complexes from temporally resolved interactions data. In fact, data already encoded into graphs are intrinsically ill-suited for the task--since they have already been ``projected'' into pairwise relations (the links of the graph). 
Although recovering the hidden higher-order interactions from pairwise networks surely represent a challenging task, recent efforts have addressed this problem with a Bayesian approach~\cite{young2020hypergraph}. Here, leveraging high-resolution proximity contact data provided by the SocioPatterns collaboration~\footnote{http://www.sociopatterns.org/datasets/}, we consider simplicial complexes representing interactions in four different social contexts: a workplace ({InVS15}~\cite{genois2015data}), a conference ({SFHH}~\cite{isella2011s}), a hospital ({LH10}~\cite{vanhems2013estimating}) and a high school ({Thiers13}~\cite{mastrandrea2015contact}). 
More precisely, as described in Ref.~\cite{iacopini2019simplicial}, we first aggregate temporally the recorded (temporal) interactions into windows of 5 minutes. Maximal cliques within each temporal window are then ``promoted'' to simplices (with associated frequency of appearance) and the final simplicial complex is formed by retaining the 20\% most frequent simplices (up to 2-simplices). More detailed information can be found in Ref.~\cite{iacopini2019simplicial}. 
The results of the stochastic simulations run on each structure are displayed in Fig.~\ref{fig:sociopatterns}, where the density of infected nodes in the stationary state is plotted as a function of $\beta_2$ for different values of $\beta_3$. Despite the very different nature of these datasets and their different generalized degree distributions, we encounter a similar phenomenology to the one described in the previous sections. 
Namely, when contributions from the higher-order interactions are stronger (higher values of $\beta_3$) we observe a lower (almost vanishing in some cases) epidemic threshold. 
Moreover, the bi-stability is present for the highest value of $\beta_3$, confirming the overall phenomenology 
obtained by the analytical approaches.

\section{Conclusions}
\label{sec:Conclusions}

In this chapter we have reviewed some recent conceptual advances in the modeling of social contagion processes, based on the idea to consider group interactions as such, and not simply as a superposition of dyadic ones. To this aim, the substrate of the contagion models has to be changed, moving from a network picture to representations by hypergraphs or simplicial complexes, which are able to represent interactions involving an arbitrary number of individuals~\cite{battiston2010structure}. 
Notably, the models of interactions themselves have to be redefined, as contagion models are traditionally defined with dyadic interactions in mind. While we have not covered all the relevant literature\footnote{For instance, the authors of 
\cite{landry2020effect} study the case in which 
$\lambda_3 <0$, i.e., an individual is less likely to adopt a trend if this trend is popular in the group, and call this
ingredient the "hipster effect"; this effect also can lead to a region of bi-stability in the phase diagram
\cite{landry2020effect}. Note that heterogeneous recovery rates \cite{darbon2019disease,de2020impact} and ``complex recovery'' rates depending on the state of the surrounding individuals have also been considered in the literature \cite{iacopini2020multilayer}.
}, we have highlighted the main approaches and results, and in particular the rich phenomenology emerging from hyperedge interactions, with co-existence of continuous and discontinuous phase transitions, bi-stability regions and 
critical mass phenomena.

Moreover, while the discovery of this rich phenomenology has already prompted a wealth of studies and brought both analytical and numerical insights, a number of interesting points remain open.

First, few analytical or mathematical results are available regarding the nature of the phase transitions: these results
have been obtained under specific approximations or for specific structures. It would be of clear interest to have more general results on the conditions (either on the structure or on the dynamical model's rules) for the 
emergence of discontinuous transitions.

Another important point regards the availability of empirical data to feed models defined on hypergraphs. Indeed, given the popularity and convenience of the network representation,
relational datasets are traditionally represented as sets of dyadic interactions and often fail to include higher-order interactions (with some exceptions, e.g. for scientific collaboration data that is easily represented as group interactions \cite{petri2018simplicial}). 
While temporally resolved data can help understand
whether cliques in an aggregated network actually correspond to group meetings or not, as discussed in Section \ref{sec:Simulations}, using dyadic data to reconstruct the actual higher-order interactions is in general far from trivial \cite{young2020hypergraph} and it seems crucial to develop new methods to this aim. 

Empirical validation of the rich phenomenology uncovered in the models remains also very challenging. On the one hand, it has been shown that complex contagion processes might become indistinguishable from simple contagion at the population level when multiple contagion processes interact 
\cite{hebert2020macroscopic}. For simple contagions taking place along networks, it is possible to infer the structure on which the process unfolds and the process' parameters  \cite{sah2018revealing}, but the generalization to higher order processes remains an open challenge. 

Validation could also come from specifically designed experiments in which the structure of the groups in which individuals interact is controlled. In the case of networks, 
controlled experiments have indeed helped discuss the role of the interaction network structure on the emergence of conventions or on the outcome of game theoretical models
 \cite{centola2015the,gracia2012heterogeneous}. For higher order structures, such experiments would also need to be carefully crafted and performed, a difficult yet promising challenge ahead.

\bibliographystyle{spphys}

\end{document}